
\input epsf                                                               %
\def\Title#1#2{\rightline{#1}\ifx\answ\bigans\nopagenumbers\pageno0\vskip1in
\else\pageno1\vskip.8in\fi \centerline{\titlefont #2}\vskip .5in}

%
%
\ifx\epsfbox\UnDeFiNeD\message{(NO epsf.tex, FIGURES WILL BE IGNORED)}
\def\figin#1{\vskip2in}
\else\message{(FIGURES WILL BE INCLUDED)}\def\figin#1{#1}
\fi
\newwrite\ffile\global\newcount\figno \global\figno=1
\def\Fig#1{Fig.~\the\figno\xdef#1{Fig.~\the\figno}\global\advance\figno
 by1}
%
%
%
%
\def\ifig#1#2#3#4{
\goodbreak\midinsert
\figin{\centerline{\epsfysize=#4truein\epsfbox{#3}}}
\narrower\narrower\noindent{\ninepoint
{\bf #1:}  #2\par}
\endinsert
}
\input jnl
\def\apm{{\alpha^\prime}}
\def\ninepoint{\def\rm{\fam0\ninerm}}

\rightline{NSF-ITP-93-58}
\rightline{hep-th/9309152}
\title HAIRY BLACK HOLES IN STRING THEORY

\author Steven B. Giddings$^1$,
\author Jeffrey A. Harvey$^2$,
\author J.G. Polchinski,
\author Stephen H. Shenker$^3$, and
\author Andrew Strominger$^1$

\affil Institute for Theoretical Physics, University of California, Santa
Barbara, CA 93106-4030

\abstract{Solutions of bosonic
string theory are constructed which correspond to four-dimensional
black holes with axionic quantum hair. The basic building blocks
are the renormalization group flows of the $CP1$ model with
a $\theta$ term and the $SU(1,1)/U(1)$ WZW coset conformal field theory.
However the solutions are also found to
have negative energy excitations, and are accordingly expected to
decay to the vacuum. }
\vskip1truein
\singlespace{\it
\noindent 1. Permanent address: Dept. of Physics, University of California,
Santa Barbara, CA 93106.
\vglue.05in
\noindent 2. Permanent address: Enrico Fermi Institute, University of Chicago,
5640 So. Ellis Ave.,
Chicago IL 60637.
\vglue.05in
\noindent 3. Permanent address: Dept. of Physics and Astronomy, Rutgers
University, Piscataway, NJ 08855-0849}

\doublespace
\vfill \eject

The classical endpoint of gravitational collapse is expected to be a simple
object: a stationary black hole characterized solely by its mass, charge and
angular momentum. This expectation is strengthened by the famous ``no-hair"
theorems [\cite{nhr}]. These theorems assume a specific field content, but they
generally suggest that the stationary, stable black hole configurations are
labeled only by conserved quantities associated with local gauge symmetries.

Quantum mechanics leads to a further balding of black holes. For example,
Hawking radiation will cause a charged black hole to radiate away mass until it
reaches the extremal (mass $=$ charge) limit. Thus one less quantum number is
required to characterize the stable endpoint.

Several years ago it was suggested that quantum mechanics might lead to hair
growth as well as hair loss [\cite{ghr}]. In theories with axion strings, an
arbitrary phase can be associated with the process of lassoing a black
hole with an axion string.  This phase arises from the axion-string
interaction Lagrangian
 $$
 S_I={T}\int_\Sigma B.\eqno(sss)
$$
The integral extends
over the string worldsheet $\Sigma,~~T$ is the string tension and $B$ is the
axion two-form potential. If $B$ is given by the closed but not exact two-form
obeying $\int_{S^2}B=\theta L^2$ (where $\theta$ is dimensionless and $L$
is the appropriate distance scale)
for any two sphere surrounding the horizon, it
follows that a string which lassos the black holes picks up a
phase $\theta  T L^2/\hbar$. $\theta $ is classically unobservable,
but may be measured quantum mechanically in
Aharonov-Bohm type interference experiments. It is thus a new
quantum number, or ``quantum hair" associated to a quantum black hole.
Generalizations of this idea involving discrete
$Z_N$  gauge symmetries were discussed in
[\cite{kw}].

Quantum hair has no perturbative effect on the Hawking radiation rate, so
large hairy black holes will evaporate as usual. However, for
small black holes, non-perturbative effects could be important and one
might suspect the existence of hairy extremal black holes[\cite{ghr}].
These objects would be stabilized against Hawking radiation by
quantum hair, just as charged extremal black holes are stabilized by their
classical electromagnetic hair.

Evidence for such extremal objects
was found (in the context of discrete gauge hair)  in
the elegant analysis of Coleman, Preskill and Wilczek [\cite{cpw}].
A Euclidean instanton which can be described as a virtual string lassoing the
black holes was shown to slow down the Hawking radiation rate,
suggesting that it might actually turn off at a critical value of the mass.
However, unlike
electromagnetic charge, axion  (or discrete) charge is periodic and cannot be
made arbitrarily large. Hairy extremal black holes are therefore typically
planckian objects, and their existence can not ordinarilly be determined from
low-energy semiclassical gravity. A complete quantum theory of
gravity - including
all higher dimension operators - is required.

In this paper we will use bosonic string theory to
demonstrate the existence of hairy extremal
black holes. In string theory, non-zero $\theta$ leads to a deformation of the
classical solutions (which vanishes exponentially at large distances)
and some exact classical
solutions will be found as conformal field theories. The
general solution will be qualitatively
described in terms of two-dimensional renormalization
group flows.  We will find a tachyonic excitation in the
spectrum, so these objects are unstable---this is in addition to the usual
tachyonic instability of string theory.  We briefly discuss the possibility
of generalization to the superstring.


Classical solutions of bosonic string theory
are provided by conformally invariant
two-dimensional sigma models. The sigma model corresponding to a
four-dimensional hairy black hole with line element
$$
ds^2=-N^2(\rho)dt^2+d\rho^2+R^2(\rho)d^2\Omega, \eqno(lmt)
$$
and axion hair
$$
B={\theta \epsilon \apm \over 2}, \eqno(bte)
$$
and dilaton field $\Phi(\rho)$
 may be written\footnote{*}{Note that, following convention, the
$\theta$ term in this action
differs from \(sss) by a factor of $\hbar$.}
$$
\eqalign{
S_\sigma & =  {1 \over \pi \apm }\int d^2\sigma \Bigg(
 -N^2(\rho)\partial_+t\partial_-t+\partial_+\rho\partial_-\rho+
{\apm }{\cal R}^{}_{+-}\Phi(\rho)\cr
&+ R^{2}(\rho)G_{\mu\nu}\partial_+X^\mu\partial_-X^\nu+{\apm
\theta \over 2}\epsilon_{\mu\nu}\partial_+
X^\mu\partial_-X^\nu\Bigg)\cr}\eqno(smn)
$$
 where $G_{\mu\nu}~~(\epsilon_{\mu \nu})~~\mu,\nu=1,2$ is
 the unit metric (volume form) on the
two-sphere, $\apm=\hbar /2 \pi T$ and $\int d^2{\sigma}{\sqrt g}
{\cal R}^{}=4\int d^2{\sigma}{\cal R}^{}_{+-}=8\pi$ on $S^2$.
We wish to find $c=4$ conformal field theories of the form \(smn).
The additional $c=22$ CFT required for $c=26$ will be suppressed.

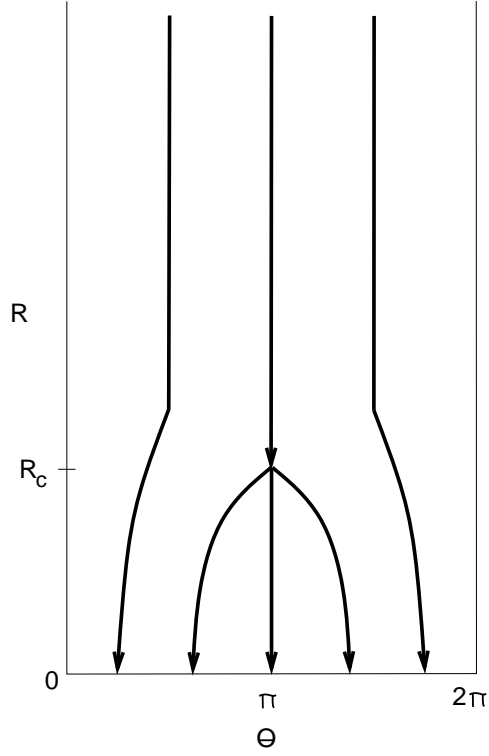
\ifig{\Fig\renormflow}{Shown is a sketch of the renormalization group flows
for the CP1 model.  An infrared fixed point lies at $\theta=\pi$, $R=R_C$.
}{hbhst.fig}{4.00}

Building blocks in our constructions are sigma models on
$S^2$ - or $CP1$ models - with $\theta $
terms, corresponding to the last two terms in \(smn) with constant $R$. These
have been the object of extensive investigations.  The
renormalization group flows in the $\theta ,~R$ planes
are depicted in figure~1.
The renormalization group flows have a fixed point at $\theta =\pi,R=R_c$
corresponding to a $c=1$ CFT, a free scalar at its
self-dual $SU(2)$-invariant radius.  This was first conjectured by Haldane
[\cite{hal}], who argued that the critical behavior of the spin-$s$
antiferromagnetic chain was governed by the $CP1$ sigma model, at $\theta =
0$ for integer spin and at $\theta=\pi$ for half-integer.  The critical
behavior for $s = \frac{1}{2}$ was already known, by bosonization and Bethe
ansatz, to be given by the scalar at the self-dual radius.
Subsequent work ([\cite{af}] and references therein; see also [\cite{fr}]
for a review)
confirmed this conjecture and filled out
the phase diagram.
The $CP1$ sigma model with $\theta$ term has been applied previously to string
theory black holes by Kogan [\cite{kogan}], with a different interpretation of
the renormalization group flow.

A $c=4$ CFT can be obtained
(following previous constructions [\cite{gs,gps}]) by simply taking the
tensor product with an $SU(1,1)/U(1)$ level $k=8,~c=3$ coset model.  By
representing this coset model as a sigma model, it was identified [\cite{wtt}]
as a two-dimensional black hole. The full sigma model action with
four-dimensional target space is, to leading order in $\apm$
$$
\eqalign{
S_\sigma &= {1 \over \pi \apm} \int d^2\sigma\Bigg[-{\rm tanh}^2
(\rho/\sqrt{6\alpha'})\,
\partial_+t\partial_-t+\partial_+\rho\partial_-\rho\cr
& + {\apm }{\cal R}^{}_{+-}(-\ln \cosh
(\rho/\sqrt{6\alpha'})\, +\Phi_h)+(R^2_cG_{\mu \nu} + {\pi\apm \over 2}
\epsilon_{\mu \nu})\partial_+X^\mu \partial_-X^\nu\Bigg],\cr}
\eqno(wxz)
$$
where $\Phi_h$ is an arbitrary constant. \(wxz) corresponds to a special type
of
four-dimensional hairy black hole. There is no asymptotically flat region; the
two-spheres have radius $R_c$ for all $\rho$ and $\theta $ is restricted to
equal
$\pi$.

The one free parameter is the constant $\Phi_h$, the value of the dilaton at
the horizon $\rho=0$, which is related to the ADM mass of the
black hole[\cite{wtt}].
At the quantum level, Hawking radiation presumably drives
the mass to zero. In this
limit the horizon moves off to infinity and the action is simply (after a
coordinate transformation)
$$
\eqalign{
S_\sigma &=  {1 \over \pi \apm}\int d^2\sigma
\Bigg[-\partial_+t\partial_-t+\partial_+\rho\partial_-\rho-
\sqrt{ \frac{\apm }{6} } {\cal
R}^{}_{+-}\rho\cr &+(R_c^2G_{\mu \nu}+ {\pi\apm \over 2} \epsilon_{\mu
\nu})\partial_+X^\mu\partial_-X^\nu\Bigg]\cr}. \eqno(xth) $$
This represents an extremal black hole
with quantum hair.

It is of interest to find hairy extremal black holes which are
asymptotically flat and thus
might exist in our universe. These can be described starting from
the renormalization group flows in figure~1. Consider a point $\theta =\pi$,
with $R$
just above the fixed point $R_c$. For small $R-R_c$, the  corresponding
sigma model is
nearly conformally invariant and has a small beta function
$$
\mu\partial_\mu\ln R(\mu)=\gamma(R-R_c)^2.\eqno(bta)
$$
where $\gamma > 0$. The approach to the fixed point is along the
perturbation $j^a_z j^a_{\bar z}$ and so is
marginal as indicated by the quadratic beta function [\cite{af}].
A conformal field theory to order $R-R_c$ can be constructed by dressing the
action with the $\rho$ field
$$
\eqalign{
S &\sim  {1 \over \pi \apm}
\int d^2\sigma\Bigg(-\partial_+t\partial_-t+\partial_+\rho
\partial_-\rho-\sqrt{\apm\over6}{\cal R}^{}_{+-} \Bigl\{ \rho +
O(\rho^{-2})
\Bigr\}\cr
&\Bigl\{ R^2_c- 2 R_c / \gamma \rho + O(\rho^{-2}) \Bigr\} G_{\mu \nu}+
{\pi\apm \over 2}\epsilon_{\mu \nu}\partial_+X^\mu\partial_-X^\nu\Bigg)\cr}
\eqno(aan)
$$
near $\rho = -\infty$.
Presumably it is possible to correct the action \(aan) to obtain a CFT in a
power series in $R-R_c={-1 / {\gamma \rho}}$ in a neighborhood of the fixed
point. As
$\rho\to-\infty$, $R$ approaches $R_c$, and the geometry
approaches the previously discussed extremal black hole. As $\rho$ increases,
the radius $R$ increases, and the black hole ``throat" begins to open up. When
$R-R_c$ is of order one the radius rapidly increases and one enters the mouth
region. At this point the expansion parameter used in constructing CFT's by
dressing renormalization group flows breaks down, and we have no quantitative
tools to analyze the theory.  However, if
one assumes that $R$ passes through the mouth region to larger values, the
theory can be analyzed in an expansion in ${1/ R}$ on the other side of
this region. To leading order in this expansion, conformal invariance and
$c=4$ implies  that $R\propto\rho$ and $\Phi$ is constant. Thus it is
plausible to assume that the theory ties on to an asymptotically flat geometry.

Other renormalization group trajectories can
similarly be used to construct extremal black holes.
For $\theta$ near but not equal to $\pi$, there is a
long throat region produced by $R$ lingering near the fixed point. In this
region a construction of the type \(aan) yields an approximate CFT. However for
$\theta \neq\pi$, $R$ eventually becomes small and the approximations break
down
   .
At the $R=0$ infrared fixed point all excitations are infinitely massive and a
spacetime interpretation of the theory is no longer possible [\cite{gps}].  If
$\theta $ is not near $\pi$, a throat region never forms, and one immediately
descends into the vicinity of the infrared fixed point.

As indicated in figure~1, the perturbation in the $\theta$ direction away from
$\theta=\pi$ is relevant.  It corresponds to the $(j,\tilde j) =
(\frac{1}{2},\frac{1}{2})$ primary, of weight $(\frac{1}{4},\frac{1}{4})$.
Expanding around a linear dilaton background,
the mass-squared (including a term
$(\nabla \Phi)^2$ from the linear dilaton),
is tachyonic, $m^2 = -17/6\apm$.  The extremal object is therefore classically
unstable, as are the near-extremal objects with long throats.   Presumably they
will decay by emission of a radial axion gradient, to
flat $R^3$ with $\theta=0$.  The would-be hair, like a bad toupee,
slips off.


The picture of the $\theta =\pi$ extremal black hole is similar to that
found for
magnetically charged extremal black holes [\cite{ghs,gps}]
which, for large charge, can be
analyzed perturbatively. There is an asymptotically flat region, and a mouth
connecting on to a semi-infinite throat region. These also share with
the present construction the feature that the time coordinate is a free field
and plays only a spectator role in the construction.
The solutions with $\theta$ neither $0$ nor $\pi$ resemble the $Q=\pm 1$
solutions of  [\cite{gps}] in the degeneration to
a massive field theory at the origin.

A qualitative - but not quantitative - picture of the
structure of hairy black holes can be developed in a mini-superspace type
approximation. The renormalization group
flows in figure~1 originate entirely from
non-perturbative worldsheet instantons which wrap around the horizon, since
those are the only configurations sensitive to $\theta$. The spacetime
effective
action which incorporates these instanton effects is non-local. However in
an $S$-wave approximation in which all configurations are required to be
spherically symmetric, a world sheet instanton is represented by a
point in the two-dimensional $\rho,~t$ plane. Summing over
spherically symmetric world sheet instantons is then equivalent
to summing over ordinary instantons in the two-dimensional effective
theory. The effects of such instantons are reproduced by adding to the
action the operator whose effects mimics that of the
instanton\footnote{$\dagger$}
{In the case at hand the single instanton action is infrared
divergent. However, as explained in [\cite{claw}], this does not mean that
instanton effects can not be summarized by a local operator.
Rather, a single insertion of the appropriate
operator must reproduce the infrared divergence.}. The result is
$$
\eqalign{
S={2\pi \over \kappa^2}\int d^2 x \sqrt{-g}e^{-2\Phi}
 \Bigg[&  R^2{\cal R}^{(2)}+2 (\nabla R)^2+2+4
R^2 (\nabla \Phi)^2\cr
&-2 \nabla^2R^2 - \frac{\apm^2}{4 R^2}(\nabla \theta)^2
+C e^{-2R^2/\apm}{\rm cos}\theta \Bigg]\cr}
\eqno(ssp)
$$
where $x =(\rho,~ t)$. C is a positive determinant.
The first six terms are obtained
by spherical reduction of the four-dimensional string action with the ansatz
$$
\eqalign{
ds^2&=g_{ab}(x)dx^a dx^b +R^2(x)d^2\Omega,\cr
B&={\apm \over 2}\theta(x)\epsilon .}
\eqno(stz)
$$
The last term in \(ssp) reproduces the effects of string instantons.
One can think about this in two ways.  The first is the usual
string sigma model point of view, where the string wrapping the black hole is
a world-sheet instanton, and the last term in the action represents the
contribution of these instantons to the beta function.
The fact that the dilute instanton approximation
suggests a $\theta=\pi$ fixed point of the $CP1$ model was noted in
[\cite{lev}].
Alternately we can think of them
- as in [\cite{cpw}] - just like ordinary
spacetime instantons involving solitonic strings wrapping
around a black hole\footnote{**}{In [\cite{cpw}] the effects of string
instantons are - in contrast to the present case - non-perturbative in $\hbar$
because they keep $T=\hbar /2 \pi \apm$ (rather than $\apm$) fixed as $\hbar
\rightarrow 0$.}.
This should be valid at large $R$ because - in the
spirit of [\cite{daha}] - the low-energy effective
field theory does not know if the core of the string contains a fundamental
string or resembles a smooth soliton.


The equations of motion following from \(ssp) have a solution with constant
$R$ and constant $\theta=\pi$.  However it is only suggestive since \(ssp) can
not be  trusted when $R$ is small and the instantons are not dilute.
This approximation also misses the fact
that the approach to the fixed point is marginal.

The reduced action \(ssp) can be thought of as
describing spherical four-dimen\-sion\-al dilaton gravity
coupled to a scalar field $\theta(x)$ with a field dependent potential that
vanishes asymptotically. Non-extremal hairy black hole solutions
can be constructed by solving the radial equations with boundary conditions
imposed at the horizons. Since the equations degenerate at the horizon,
there are (after gauge fixing) only three independent initial data
which may be taken to be the horizon values
$\theta_h$, $\Phi_h$ and $R_h$ of $\theta$,
$\Phi$ and $R$.
$R_h$ directly determines the horizon area,
while $\Phi_h$ determines the asymptotic value of $\Phi$.
If $\theta$ were an ordinary massive field there would be
two possible asymptotic
solutions: one which grows and one which decays exponentially.
Unless one takes $\theta_h=0$, there would
be some admixture of the growing solution and the spacetime would not be
asymptotically flat. This is in  accord with the no hair theorems for scalar
fields. On the other hand if $\theta$  were exactly massless the asymptotic
solutions go as a constant plus $1/R$,  and there is no danger of destroying
asymptotic flatness. In fact the solution will have $\theta=\theta_h$
everywhere. The action~\(ssp) is somewhere between the  massive and massless
case. Because the mass vanishes asymptotically,  there are no  growing modes
and no fear of destroying asymptotic flatness. On the  other hand since the
potential is non-zero $\theta$ will not be  constant if $\theta_h \neq 0,\pi$.
In general the phase measured in  string interference will be given by the
asymptotic value of $\theta(x)$ which is  non-trivially related to $\theta_h$.


It is natural to attempt a similar construction for the superstring. In this
case, however, $\theta$ can be
eliminated by a chiral fermion rotation so there is
no analog of figure~1.  The analogous effect for black holes with discrete
gauge hair was discussed in [\cite{gh}]. However for a collection of neutral
black holes
characterized by different values of $\theta$, there is no globally
defined chiral rotation which eliminates all the $\theta$s. Thus we
do expect quantum hair to arise, although the conformal field theoretic methods
described herein are inadequate to describe it.

In closing we wish to discuss the observability of the axion hair in this
solution.  There is a limit, albeit artificial, in which it can be measured.
This is the limit of string tree level, where we are doing conformal field
theory in a fixed background.  This limit is partly classical and partly
quantum mechanical.  The background fields do not fluctuate, but the
propagation of test strings is quantum mechanical; for example, the vertex
operators satisfy wave equations.  A spherical world-sheet can loop
the black hole, so interference effects from $\theta$ will appear in
tree-level amplitudes.  Note also that because the background is classical
the field satisfies $<{\Delta B}>^2=<{(\Delta  B)^2}>$.  The left-hand side
here is of two-instanton order.  The right-hand side has both one- and two-
instanton contributions, but the first enters at string loop order
and so is suppressed in the limit discussed here.

Quantum effects make the axion hair difficult to measure for several
reasons. In general $\theta$ eigenstates  are not
energy eigenstates, and so the $\theta$-mode of the axion field will rapidly
fluctuate (we believe this is equivalent to the arguments of [\cite{cpw}]).
In the present case this is exacerbated by the presence of the tachyonic mode.

In conclusion, under closer inspection axion hair turns out to be a toupee
in bosonic string theory: it does not provide a new quantum label for
stable, extremal black holes. It
remains a logical and interesting possibility that genuine
quantum hair
could exist in other contexts, such as superstring theories
or discrete gauge symmetries.

\head{Acknowledgements}

This work was supported in part by the National
Science Foundation under Grants
No. PHY89-04035,  PHY91-16964, PHY90-00386 and
by Department of Energy grants DOE-91ER40618 and DE-FG05-90ER40559,
and by NSF PYI grant PHY-9157463 to SBG and NSF PYI grant
PHY-9196117 to JAH. We are grateful to
J. Friedman, G. Horowitz and N. Seiberg for discussions, to
the kindly staff of the UCEN dining services for their hospitality,
and to the  excellent taco special for inspiration.

\head{References}

\refis{nhr} V.~Ginzburg and L.~Ozernoi, {\sl Soviet Phys., JETP\/} {\bf 20}
(1965), 689;
 A.~Doroshkevich, Ya.~Zeldovich and I.~Novikov, {\sl Soviet Phys., JETP\/}
{\bf 22} (1966), 122; W.~Israel, {\sl Phys. Rev.\/} {\bf 164} (1967), 1776;,
{\sl Comm. Math. Phys.\/} {\bf 8} (1968), 245; B.~Carter, {\sl Phys. Rev.
Lett.\/} {\bf 26} (1971), 331; J.~D.~Bekenstein, {\sl Phys. Rev.\/} {\bf D5}
(1972), 1239; J.~B.~Hartle, in {\it Magic\ without\ Magic,\/} ed. J.~Klauder
(Freeman, 1972); C.~Teitelboim, {\sl Phys. Rev.\/} {\bf D5} (1972), 2941;
 R.~H.~Price, {\sl Phys. Rev.\/} {\bf D5} (1972), 2419: {\it ibid.\/} 2439.

\refis{wtt} E.~Witten, {\sl  Phys. Rev.\/} {\bf D44} (1991), 314.

\refis{gs} S.B.~Giddings and A.~Strominger,
{\sl Phys. Rev. Lett.\/} {\bf 67} (1991), 2930.

\refis{hal} F. D. M. Haldane, {\sl Phys. Lett. \/} {\bf 93A} (1983), 464.

\refis{af} I. Affleck in {\sl Fields, Strings and Critical
Phenomena,}
eds. E. Brezin and Z. Zinn-Justin, North Holland (New York) 1990.

\refis{fr} E. Fradkin, {\sl Field Theories of Condensed Matter Systems,}
Addison-Wesley (Redwood City) 1991.

\refis{kogan} I. Kogan, {\sl Mod. Phys. Lett.\/} {\bf A6} (1991), 3297.

\refis{ghr} M.~Bowick, S.B.~Giddings, J.~Harvey, G.~Horowitz, and
A.~Strominger,
{\sl Phys. Rev. Lett.\/} {\bf 61} (1988), 2823.

\refis{kw} L.~Krauss and F.~Wilczek, {\sl Phys. Rev. Lett.\/} {\bf 62} (1989),
1221.

\refis{cpw} S.~Coleman, J.~Preskill, and F.~Wilczek, {\sl Phys. Rev. Lett.\/}
{\bf 67} (1991), 1975;  {\sl Nucl. Phys.\/} {\bf B378} (1992), 175.

\refis{ghs}G. Gibbons and K. Maeda, {\sl Nucl. Phys.} {\bf B298} (1988) 741;
D. Garfinkle, G.~Horowitz, and A.~Strominger. {\sl Phys. Rev.\/}
{\bf D43} (1991), 3140; Erratum: {\sl Phys. Rev.\/} {\bf D45} (1992), 3888.

\refis{kwd} B. S. Kay and R. M. Wald {\sl Phys. Reports} {\bf 207(2)} (1991).

\refis{lev}H. Levine, S. B. Libby, A. M. M. Pruisken {\sl Nucl. Phys.}
{\bf B240} (1984), 71; V. Knizhnik and A. Morozov, {\sl JETP Lett.\/} {\bf 39}
(1984), 240.

\refis{gps} S.B. Giddings, J. Polchinski and A. Strominger,
``{Four Dimensional Black Holes in String Theory}'' preprint
NSF-ITP-93-62, hep-th/9305083.

\refis{claw} L. F. Abbott and M. B. Wise, {\sl Nucl. Phys.}
{\bf B325} (1989), 687; S. Coleman and K. Lee  {\sl Nucl. Phys.}
{\bf B329} (1990), 387.

\refis{daha}A. Dabholkar and J. A. Harvey, {\sl Phys. Rev. Lett.\/}
{\bf 63} (1989), 719.

\refis{gh}R. Gregory and J. A. Harvey, {\sl Phys. Rev. \/} {\bf D46} (1992)
3302.

\endreferences

\end